\documentclass[10pt]{article}

\usepackage{natbib}
\usepackage{graphicx}
\usepackage{txfonts}
\usepackage[normalem]{ulem}
\usepackage{color}
\bibpunct{(}{)}{;}{a}{}{,}

\newcommand{\qm}[1]{``#1''}

\def\sss{\scriptscriptstyle}
\def\U{{\sss \,\mathrm{U}}}
\def\L{{\sss \,\mathrm{L}}}

\def\nuL{\nu_\L}
\def\nuU{\nu_\U}

\def\4u1636{\mbox{4U~1636$-$53}}

\definecolor{gray}{rgb}{.6,.6,.6}
\definecolor{green}{rgb}{0,.6,0}
\definecolor{red}{rgb}{0,0,.6}
\newcommand*{\cancel}[1]{\textcolor{gray}{}}

\begin{document}
\begin{center}
\noindent
{{\bf\Large Distribution of kilohertz QPO frequencies and their ratios in the atoll source 4U~1636-53}}
\end{center}
\vspace{1ex}
\noindent
 {Gabriel T\"or\"ok$^{1}$,
  Marek A. Abramowicz$^{1,2,3}$,
  Pavel Bakala$^{1}$,
  Michal Bursa$^{4}$,
  Ji\v{r}\'{\i} Hor\'{a}k$^{4}$,
  W{\l}odek Klu\'zniak$^{3,5}$,
  Paola Rebusco$^{6,7}$,
  Zden\v{e}k Stuchl\'{\i}k$^{1}$
  }
\\\\
{\footnotesize{
  $^{1}$ Institute of Physics, Faculty of Philosophy and Science, Silesian
  University in Opava, Bezru\v{c}ovo n\'{a}m. 13,746-01 Opava, CZ
  {\\~~\scriptsize e-mail: terek@volny.cz, pavel.bakala@ fpf.slu.cz, zdenek.stuchlik@fpf.slu.cz }
  \\\\
  $^{2}$ Department of Physics, G\"oteborg University, S-412 96 G\"oteborg, SE
  {\\~~\scriptsize e-mail: marek.abramowicz@physics.gu.se}
  \\\\
  $^{3}$ Copernicus Astronomical Centre PAN, Bartycka 18, 00-716 Warsaw, PL
  {\\~~\scriptsize e-mail: arktur@camk.edu.pl, wlodek@camk.edu.pl}
 \\\\
  $^{4}$ Astronomical Institute of the Academy of Sciences, Bo\v{c}n\'{\i}~II 1401/1a, 141-31 Praha~4, CZ
  {\\~~\scriptsize e-mail: bursa@astro.cas.cz, horak@astro.cas.cz}
  \\\\
  $^{5}$ Johannes Kepler Institute of Astronomy, Zielona Gora University, ul. Lubuska 2, 65-265 Zielona G\'ora, PL 
   \\\\
  $^{6}$ MIT Kavli Institute for Astrophysics and Space Research, 77 Massachusetts Avenue, 37, Cambridge, MA 02139, US {\\~~\scriptsize e-mail: pao@space.mit.edu}
  \\\\
   $^{7}$ Max-Planck-Institute for Astrophysics, Karl-Schwarzschild-Str. 1,
D-85741 Garching, D
  }}
%
\\\\
{\footnotesize {\bf{Abstract.}} A recently published study on long term evolution of the frequencies of the kilohertz quasi-periodic oscillations (QPOs) in the atoll source 4U 1636-53 concluded that there is no preferred frequency ratio in a distribution of twin QPOs that was \emph{inferred} from the distribution of a single frequency alone.
However, we find that the distribution of the ratio of \emph{actually observed} pairs of kHz QPO frequencies is peaked close to the $3/2$ value, and possibly also close to the $5/4$ ratio. To resolve the apparent contradiction between the two studies, we examine in detail the frequency
distributions of the lower kHz QPO and the upper kHz QPO detected in our data set.
We demonstrate that {for each of the two kHz QPOs (the lower or the upper), the  frequency distribution in all detections of a QPO} differs from the distribution of frequency of the same QPO in the subset of observations where both the kHz QPOs are detected.
We conclude that detections of individual QPOs alone should not be used for calculation of the distribution of the frequency ratios.}
\\\\{\footnotesize{{\bf Keywords:}{~X-rays:binaries --- Stars:neutron --- Accretion, accretion disks}}



\section{Introduction}
\label{section:introduction}

\cite{abr-klu:2000} suggested that the kHz twin peak QPOs, observed in the Fourier power  spectra (PDS) from accreting neutron stars, originate in a non-linear resonance that is possible only in strong
gravity. It was reported later \citep{abr-etal:2003} that the ratio $\nuL/\nuU$ of the upper and lower QPO frequency in neutron stars usually clusters close to  the rational ratio $2/3$, with some frequency pairs possibly clustering close to other ratios, such as 0.78.

\cite{bel-etal:2005} re-examined the study of \cite{abr-etal:2003} for a larger set of detections of a single kHz QPO and, on the assumption of a correlation between the observed QPO and the unobserved second QPO, confirmed that their (inverse) frequency ratio $\nuU/\nuL$ would cluster most often close to the $3/2$ value and less often close to other rational numbers (e.g., 5/4 and 4/3). Because a  distribution of the ratios of two correlated quantities is largely determined by the distribution of either one of them, \cite{bel-etal:2005} argued that the peaks in the distribution of kHz frequency ratios reported by \cite{abr-etal:2003} reflect peaks of unknown origin in the distribution of a single (upper or lower) kHz QPO. {Further,}  they argued that such clustering does not provide any useful information about a possible underlying physical mechanism.

A more recent study of \citet{bel-etal:2007b} is based on a long term evolution of the QPO  frequencies in the atoll source \4u1636 over an eighteen month period and on the results of their previous research. The authors now conclude that in fact there are no peaks in the frequency  distribution of the lower kHz QPO in this source. In keeping with their previous argument, they conclude that there are no peaks in the frequency ratio distribution either.

While \cite{abr-etal:2003} examine the frequency ratios in pairs of observed frequencies, both of the cited papers of Belloni et al.\ study primarily distributions of the frequencies and focus mainly on the lower QPO. Note that in most observations of \cite{bel-etal:2007b} only single QPO frequencies
 have been detected.

\begin{figure*}[t]
\vspace{1em}
\includegraphics[width=1\textwidth]{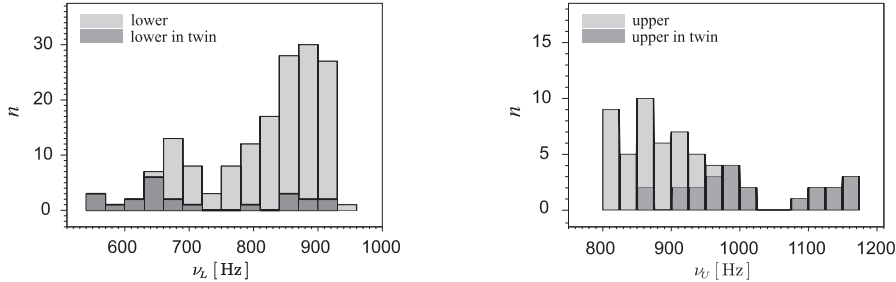}
\caption{\footnotesize {\sl Left}: Frequency histogram for the lower QPO. {The subset of lower QPOs detected simultaneously with the upper QPO is denoted by darker bars, which are labeled \qm{lower in twin}.} 
{\sl Right}: Analogous histograms for the upper QPO.}
\label{fig:distributions}
\end{figure*}


\section{Ratio vs.\ frequency distribution in the atoll source \4u1636}

Using \4u1636 data from the analysis of \cite{bar-etal:2005b}, we prepare a histogram of detections of the lower QPO over nine years (from 1996 till 2005) of monitoring by \emph{RXTE} (Figure~\ref{fig:distributions}, left), {as well as of the upper QPO (Figure~\ref{fig:distributions}, right). We restrict our study to frequencies $>400\,$Hz for the lower QPO, and $>800\,$Hz for the upper QPO.} The inaccuracies caused by the long-term frequency drift inside of continuous data segments are not important for the purposes of our paper.

The data of \cite{bar-etal:2005b} have been obtained through a shift-add procedure  carried out on individual continuous segments  of observation. In this approach each continuous data segment (corresponding to few tens minutes of an effective subset from 1.5 hour RXTE orbital period) is divided into N intervals, and searched for a QPO. The shortest usable integration time is estimated such that the QPO is detected above a certain significance in at least 80$\%$ of the N intervals; a linear interpolation is used to estimate the QPO frequency in the remaining intervals. The N PDS are frequency-shifted to the mean QPO frequency over the segment and averaged \citep{men-etal:1998}. The resulting averaged PDS representing the complete continuous segment is then fitted with one or two Lorentzians plus a constant corresponding to the counting-statistics noise level \citep{bar-etal:2005b}. 

When only one significant peak is detected, the QPO is identified as upper or lower from the parameters of the Lorentzian {and we refer to such peaks as \emph{single} QPOs. We stress that the value of the QPO frequency itself is not used to distinguish between the upper and lower QPOs---in principle, a QPO of a given frequency could be either the upper or the lower QPO}. For instance, the quality factor for the lower kHz QPO is a well determined function of the frequency, and a different function of frequency for the upper QPO, and we can use this and other relationships to identify a QPO of given frequency \citep[see][for details]{bar-etal:2005a,bar-etal:2005b,bar-etal:2005c,bar-etal:2006}.\footnote{This way of QPO identification differs from the method based on hardness diagram applied in \citet{bel-etal:2005,bel-etal:2007b}. The two methods have a different range of applicability but give comparable results {\citep[see e.g.,\ ][]{bar-etal:2005b, bar-etal:2005c}}.} {Of course, when two significant kHz QPOs are detected, the upper QPO is the one with the larger frequency, by definition.}

Consequently, the frequency values are averaged through intervals of predetermined length $\sim\!2000\,$s. \cite{bel-etal:2005,bel-etal:2007b} analyzed segments of different lengths, resulting in a larger number of detections. Hence the histograms we use here are not comparable in details to those of \citeauthor{bel-etal:2005}, even for the same RXTE data.

We take into account only the detections of oscillations with quality factor (defined as the QPO centroid frequency over the full-width of the peak at its half-maximum) $Q\,\geq\,2$ and significance (defined as the integral of the Lorentzian fitting the peak in PDS divided by its error) $S\,\geq\,3$. 

\subsection{Different distributions}

Each of the histograms in the Figure \ref{fig:distributions} clearly reveals an accumulation of frequencies in the \mbox{$\nu\sim900\,\mathrm{Hz}$} vicinity of the power spectrum. However, for the lower kHz QPO this range \mbox{$(\nuL\sim900\mathrm{Hz})$} is in the high-frequency part of  the frequency distribution of this QPO, while most detections of $\nuU$ are accumulated in the low-frequency part of its own range of variation.

If one is interested (for whatever reason) in the distribution of the frequency ratio $\nuU/\nuL$, then those observations in which both QPO peaks are simultaneously detected should be considered. Accordingly, we apply our selection criteria to \emph{simultaneous significant detections of both QPO frequencies} as well.  The histograms of the upper QPO frequency in this sample (darker bars in Figure~\ref{fig:distributions}, right panel) are strikingly different from the previous histogram of significant detections of the upper QPO (bars of lighter shade in the same figure). A new cluster of frequencies appears, in the range $\sim 1100\,$ Hz to $\sim 1200\,$ Hz, at the expense of frequencies below 900 Hz, whose occurrence is greatly diminished. 

While there is a positive correlation between the QPO frequencies (e.g., \citeauthor{abr-etal:2005:rag} 2005; see also \citeauthor{bel-etal:2005} 2005, \citeauthor{zha-etal:2006} 2006), 
\begin{equation}
\label{eq}
\nuU\approx 0.7\nuL + 520\,\mathrm{Hz},
\end{equation}
very clearly the examined data do not support the \emph{assumption} of \cite{bel-etal:2005,bel-etal:2007b} that the distribution of the ratio of two linearly correlated frequencies is determined by the distribution of one of the frequencies even when the second frequency remains undetected---there is apparently no direct link
between  the histogram of all the lower QPO detections (Figure~\ref{fig:distributions},  {\sl left}; lighter) and the histogram of the same QPO taken from the subset of twin peak QPO detections (the same figure; darker). This result should have an impact on the theory of QPOs. Although a full discussion is beyond the scope of this paper, we note that the change in the frequency distribution when a second QPO is detected may be suggestive of a physical mechanism, such as mode-coupling.

To quantify this effect we plot  the cumulative distributions of the lower and of the upper QPO, which are shown in the Figure~\ref{fig:cumulative}. Using the Kolmogorov-Smirnov (K-S) test we compare the frequency distributions of each (the upper and the lower) QPO measured in all detections,  with those measured for the same QPO when both the upper and the lower QPO are detected. We obtained the K-S probabilities $p_{L,\mathrm{KS}}=2.35\times10^{-5}$ and 
$p_{U,\mathrm{KS}}=2.24\times10^{-3}$ in the case of the lower and upper QPO respectively. Indeed, the two distributions are different in both cases. We directly conclude that detections of individual QPOs alone cannot be used for calculation of the distribution of the frequency ratios.

\begin{figure*}[t]
\vspace{1em}
\includegraphics[width=1\textwidth]{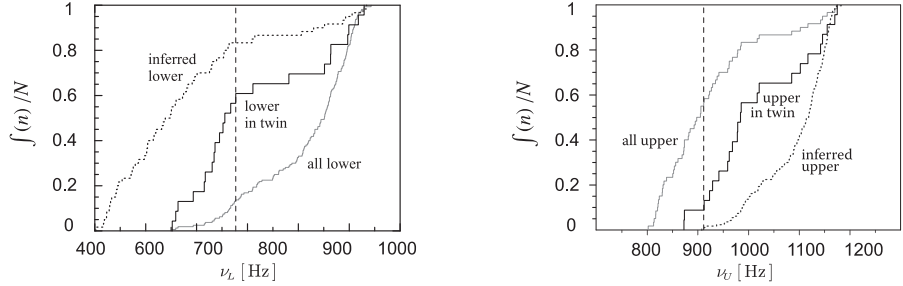}
\caption{\footnotesize {The cumulative distributions for the kHz QPOs corresponding to Fig.~\ref{fig:distributions}. {\sl Left}: Solid curves correspond to the detected lower QPOs (see left panel of Fig.~\ref{fig:distributions}), the dotted line labeled \qm{inferred lower} indicates the lower QPO frequency calculated from eq.~(\ref{eq}) using all detections of the upper QPO. The dashed vertical line shows the greatest difference $D_{\rm max}\!=\!0.515$ between the distributions of \qm{all lower} and \qm{lower in twin}. {\sl Right}: Analogous lines for the upper QPO. The dashed vertical line on right panel corresponds to the maximal difference $D_{\rm max}\!=\!0.4364$ between the \qm{all upper} and \qm{upper in twin} distributions.}}
\label{fig:cumulative}
\end{figure*}
\begin{figure*}[t]
\vspace{1em}
\begin{center}
\includegraphics[width=.48\textwidth]{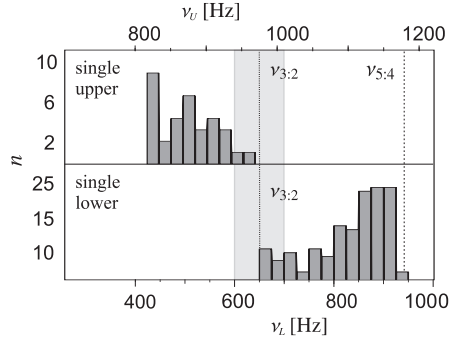}
\end{center}
\caption{\footnotesize {Distribution of single QPOs. Frequency axes are aligned according to the correlation of eq.~(\ref{eq}). The shadow denotes a 50 Hz scatter about the lower QPO frequency of 650 Hz, corresponding to a 3:2 ratio.}}
\label{fig:ratio-distr}
\end{figure*}

{It is interesting to note that the single upper QPOs are mostly detected at relatively low frequencies, while the single lower QPOs are detected at relatively high frequencies. Taking into account the linear correlation among QPO frequencies, the distributions of single lower and upper QPOs appear to be complementary, in the following sense. The lowest-frequency detection of the single lower QPO is at $\nuL = 651\,\mathrm{Hz}$, which in the linear correlation corresponds to $\nuU = 976\,\mathrm{Hz}$ while the highest-frequency detection of the single upper QPO is at $\nuU = 961\,\mathrm{Hz}$, which corresponds to $\nuL = 628~\mathrm{Hz}$. In other words, if one assumed that each of the single upper QPOs is accompanied by a lower QPO of frequency determined from eq.~(\ref{eq}), the resulting points would all fall to the left of the 3:2 line in Fig.~\ref{fig:ratios} (left panel), and if the same procedure were applied to the single lower QPOs, the resulting points would fall to the right of the 3:2 line in the same figure. This is illustrated in Fig.~\ref{fig:ratio-distr}.
Given this fact, it is not surprising that the distribution of actually detected upper (or lower) kHz QPOs is completely different from the distribution that would be predicted on eq.~(\ref{eq}) from the distribution of the other kHz QPO, when detections of single QPOs dominate the data set (Fig.~\ref{fig:cumulative}).}

\begin{figure*}[t]
\vspace{1em}
\includegraphics[width=1\textwidth]{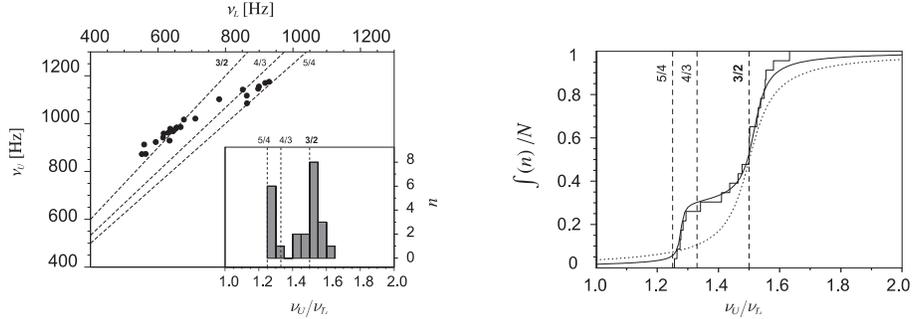}
\caption{\footnotesize {\sl Left}: {The frequencies of detected twin QPOs}. The inset shows a {corresponding} histogram of the frequency ratio. 
{\sl Right}: Cumulative distribution of the frequency ratios, the thick solid line denotes the best fit by a sum of two Lorentzians. Fit by a single Lorenzian is marked by dotted line.}
\label{fig:ratios}
\end{figure*}

\medskip

\noindent
\subsection{Possible peaks in the ratio distribution}

The left panel of Figure~\ref{fig:ratios} depicts the {mutual dependence of frequencies of the lower and upper QPO when} both were significantly detected. It also display a corresponding histogram of the frequency ratio.  As for Sco X-1 \citep{abr-etal:2003}, this histogram is peaked close to the 3/2 value, and is suggestive of the existence of a second peak.

We have fitted the distribution of the frequency ratios by the sum of two suitably normalized Lorentzians,
\begin{equation}
  p_2(r) = f\,\frac{\lambda_1/\pi}{(r-r_1)^2+\lambda_1^2} + 
    (1-f)\,\frac{\lambda_2/\pi}{(r-r_2)^2+\lambda_2^2},
\end{equation}
where $r=\nuU/\nuL$ is the frequency ratio and $r_1$, $r_2$, $\lambda_1$, $\lambda_2$ and $f$ are free parameters. Their values obtained by the maximum likelihood method are $r_1 = 1.52$, $r_2=1.28$, 
$\lambda_1=0.0327$, $\lambda_2=0.0913$ and $f=0.722$, reaching K-S probability $p_{2,\mathrm{KS}}=0.918$. The best fit by a single Lorentzian, with $r_0=1.50$ and $\lambda_0=0.0597$ (dotted line in Figure~\ref{fig:ratios}) gives the K-S probability $p_{1,\mathrm{KS}}=0.340$. Both fits are acceptable. In the right panel of Figure \ref{fig:ratios} we compare cumulative distributions of the observed frequencies with both double and single Lorentzians.


\section{Conclusions}

We have demonstrated for a set of uniform data \citep{bar-etal:2005b} that the frequency distribution of a single kHz QPO is not equivalent to the distribution of the corresponding frequency when a pair of kHz QPOs have been detected.

We stress that if there is a one-to-one correspondence between the frequencies and their ratio, as is the case for linear functions with a non-vanishing intercept, the question whether to consider the QPO frequency distribution or the ratio distribution as fundamental is one of theoretical assumptions, as the two distributions are mathematically equivalent. 
{However, the distribution of a single kHz QPO frequency is not predictive of the distribution of two frequencies detected simultaneously, nor of the distribution of their ratio, even if these frequencies are correlated when both are actually detected. Thus, the study of \citet{bel-etal:2007b}, who conclude that ``\emph{there is no preferred frequency or frequency ratio in \4u1636}'' is based on an invalid assumption, and cannot be accepted as applying to the distribution of ratios, as long as it is based on the detection of a single frequency.}

{The finding that the frequency distribution of a QPO depends on whether or not a second QPO can be detected as well} should restrict models of the physical origin of the QPO and of X-ray flux modulation,
regardless of whether or not the value of the frequency ratio is clustered about the specific value of 3/2. 
\bigskip

%
\noindent
{\footnotesize{\bf{Acknowledgements.}}
~We thank Didier Barret for providing the data and software on which this paper builds and for several discussions. We have also benefited from helpful comments by Tomek Bulik. We thank the referee for very useful suggestions. The authors are supported by the Czech grants MSM~4781305903 and LC06014, by the Polish grants KBN N203~009~31/1466 and 1P03D~005~30.}


\end{document}